\documentclass[twocolumn]{aastex631}

\usepackage{amsmath}
\usepackage{graphicx}
\usepackage{longtable}
\usepackage{appendix}
\definecolor{myblue}{RGB}{0, 160, 240} 
\definecolor{mygreen}{RGB}{0, 180, 0}

\newcommand{\new}{\textcolor{black}}
\newcommand{\newc}{\textcolor{black}} %purple

\hypersetup{colorlinks = true, 
citecolor = {blue},  % color of references in the text
linkcolor={blue}, 
urlcolor={blue}}

\shorttitle{A Re-analysis of the First Cepheid Period-Luminosity Relation}
\shortauthors{Breuval L., Huang C. D. \& Riess A. G.}

\begin{document}

\title{The Legacy of Henrietta Leavitt: \\ A Re-analysis of the First Cepheid Period-Luminosity Relation}

\author[0000-0003-3889-7709]{Louise Breuval}
\altaffiliation{ESA Research Fellow}
\affiliation{European Space Agency (ESA), ESA Office, Space Telescope Science Institute, \\ 3700 San Martin Drive, Baltimore, MD 21218, USA}
\email{lbreuval@stsci.edu}

\author[0000-0001-6169-8586]{Caroline D. Huang}
\altaffiliation{NSF Astronomy and Astrophysics Postdoctoral Fellow}
\affiliation{Center for Astrophysics $\vert$ Harvard \& Smithsonian, 60 Garden Street, Cambridge, MA 02138, USA}

\author[0000-0002-6124-1196]{Adam G. Riess}
\affiliation{Space Telescope Science Institute, 3700 San Martin Drive, Baltimore, MD 21218, USA}
\affiliation{Department of Physics and Astronomy, Johns Hopkins University, Baltimore, MD 21218, USA}

\begin{abstract}
%Henrietta Leavitt's discovery of the Period-Luminosity relation of pulsating stars appeared in 1912 in the Harvard College Observatory Circular. Her conclusion – that Cepheids could potentially serve as cosmic yardsticks – was based on a sample of 25 variable stars in the Small Magellanic Cloud, for which Leavitt obtained light curves and mean magnitudes from visual inspection of photographic plates. This paper opened the way to the first precise series of distance measurements and, a few years later, to the discovery of the expansion of the Universe by Edwin Hubble. We analyze and reproduce Leavitt's first Period-Luminosity (P-L) relation using modern data and methods of Cepheid classification and analysis. Using only data recorded in Leavitt's notebooks, we assess the quality of her results using her light curves, her measured periods, and the slope and scatter of her Period-Luminosity relations. We show that modern methods, for the same objects, reduce the P-L scatter by a factor of two. Furthermore, we report unusual pulsational behavior from BZ Tuc (HV 821), which we were motivated to re-examine due to the lack of consistency between Leavitt's and modern observations for this object. We find that overall, Leavitt’s results are in excellent agreement with our contemporary measurements, a testament to the visionary quality of her work. 

Henrietta Swan Leavitt’s discovery of the relationship between the period and luminosity (hereafter the Leavitt Law) of 25 variable stars in the Small Magellanic Cloud, published in 1912, revolutionized cosmology. These variables, eventually identified as Cepheids, became the first known ‘standard candles’ for measuring extragalactic distances and remain the gold standard for this task today. Leavitt measured light curves, periods, and minimum and maximum magnitudes from painstaking visual inspection of photographic plates. Her work paved the way for the first precise series of distance measurements that helped set the scale of the Universe, and later the discovery of its expansion by Edwin Hubble in 1929. Here, we re-analyze Leavitt’s first Period-Luminosity relation using observations of the same set of stars but with modern data and methods of Cepheid analysis. Using only data from Leavitt’s notebooks, we assess the quality of her light curves, measured periods, and the slope and scatter of her Period-Luminosity relations. We show that modern data and methods, for the same objects, reduce the scatter of the Period-Luminosity relation by a factor of two. We also find a bias brightward at the short period end, due to the non-linearity of the plates and environmental crowding.  Overall, Leavitt’s results are in excellent agreement with contemporary measurements, reinforcing the value of Cepheids in cosmology today, a testament to the enduring quality of her work. \\
\vspace{0.5cm}
\end{abstract}

\section{Introduction} 

\begin{figure*}[t!]
\centering
\includegraphics[height=8.0cm]{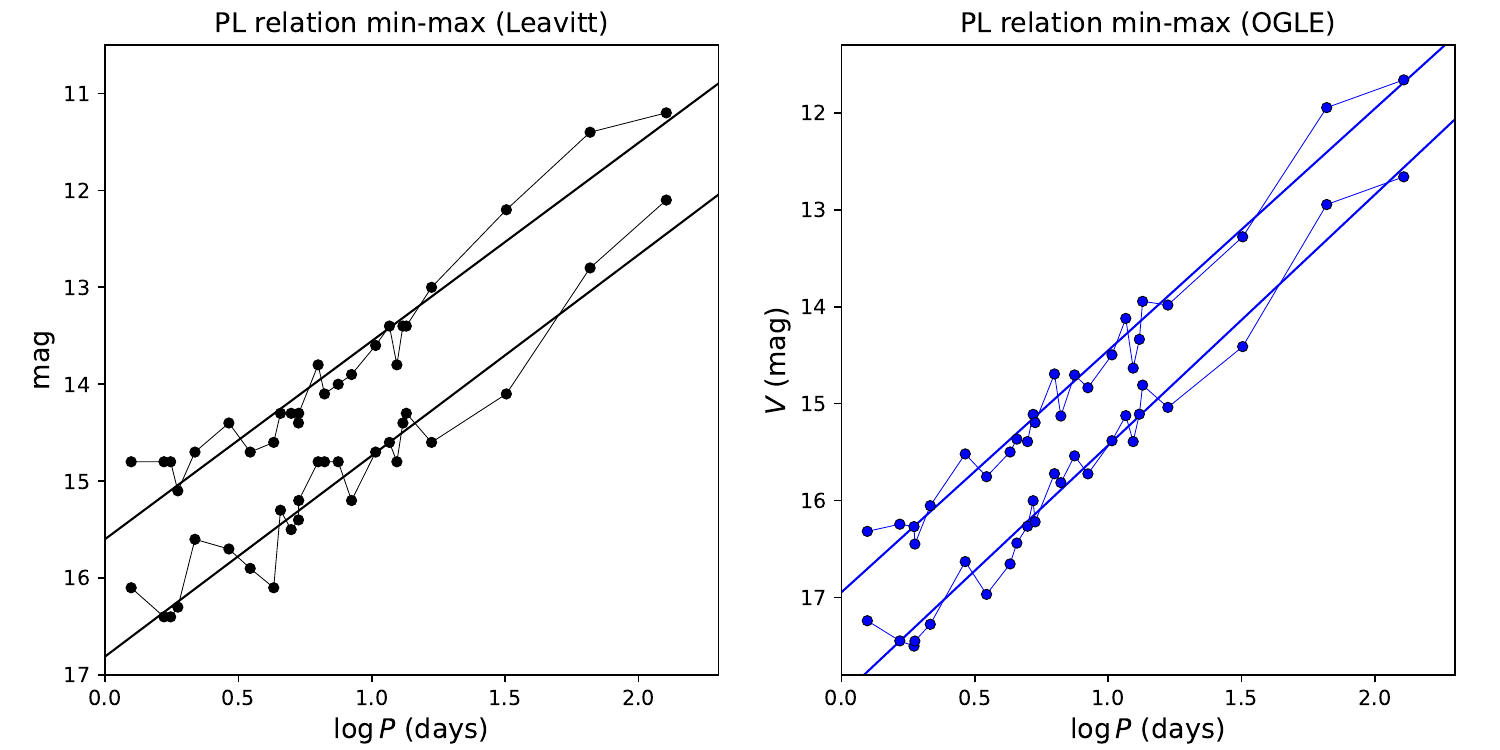} 
\caption{\textbf{Left}: Original Period-Luminosity relation (minimum and maximum light) published by \citet{Leavitt1912} in a "provisional" scale of magnitudes. Data are taken from their Table 1. \textbf{Right}: Same as left with pulsation periods and apparent magnitudes in the $V$-band from the OGLE survey \citep{Soszynski2015}, uncorrected for reddening. \new{The thin lines between data points do not represent any physical relation but reproduce the style of Leavitt's first plot of the P-L relation \citep{Leavitt1912}.} \\}
\label{fig:PL_Leavitt}
\end{figure*}

%or should this say something like Historical Context? - CH

Distances play a crucial role in astronomy, as they establish the absolute scales for a wide range of astronomical objects.  While direct geometric distances can be obtained to the nearest objects ($\sim$5 kpc with \emph{Gaia}), determining distances to extragalactic systems typically necessitates the use of a cosmic distance ladder, where multiple distance measurement techniques are cross-calibrated over the range of distances for which they overlap. Most distance ladders use geometric distances as the first ``rung” to calibrate the absolute luminosity of a standard candle – a bright astrophysical source with a known intrinsic luminosity, which can then measure systems out of the range of geometric methods \citep[e.g.][]{Riess2022}. 

Cepheid variables, the first standard candles to be developed, were noticed by Henrietta Swan Leavitt in her seminal paper cataloguing 1777 variables in the Magellanic Clouds \citep{Leavitt1907}. She remarked that for a subset of Small Magellanic Cloud (SMC) variables, \emph{``It is worthy of notice that ... the brighter variables have the longer periods.  It is also noticeable that those having the longest periods appear to be as regular in their variations as those which pass their changes in a day or two."} The relation between Cepheid luminosities and periods (Period-Luminosity Relation, hereafter P-L relation or Leavitt Law) was further established in \cite{Leavitt1912} where the periods and magnitudes of 25 SMC Cepheids were reported. Only a few years later, in 1917, Sir Arthur Eddington proposed that this relationship could be driven by fundamental-mode pulsation resulting from the kappa-opacity mechanism, giving this empirical relationship a firm physical explanation \citep{Eddington_1917}. The physical size and temperature changes experienced by radially pulsating Cepheids result in periodic variations in their luminosities. By observing the stellar light curves, we can determine the periods of these variations, independent of their distance. Once the absolute scale of the Leavitt Law is calibrated using the geometric distance to the nearest Cepheids, the apparent brightnesses of more distant Cepheids can then be used to infer their true distances. 

Unfortunately Leavitt, who died in 1921, did not live to see the tremendous impact of her work. The realization that Cepheids could be used to measure distances to a variety of astrophysical systems would change our cosmological paradigm. In 1924, this led directly to Hubble's transformative discovery of the extragalactic universe. By observing Cepheids in the Andromeda Galaxy (M31) and measuring their distances, he provided definitive evidence that the extragalactic `nebulae' were not part of the Milky Way, but instead were `island Universes' comparable to our own Galaxy. Only five years later, in 1929, he used Cepheids as the foundation of his distance ladder to show that the more distant galaxies were moving away more quickly  -- evidence that convinced the scientific community that the universe was expanding \citep{Hubble1929}. 

Hubble's contributions to cosmology have been revolutionary, but arguably no less important is the fundamental discovery made by Leavitt that enabled his work. However, unlike Hubble's work, which has previously been re-examined and reconstructed \citep{Kirshner2003}, comparatively little effort has been placed in re-examining Leavitt's Cepheids and comparing her work with modern methods. Here, we re-examine the Small Magellanic Cloud Cepheids in \cite{Leavitt1912} using modern datasets and methods and compare with Leavitt's results. Furthermore, we show that Leavitt's work is remarkably consistent with contemporary observations for the same sample.

First, in Sect. \ref{sect:identification}, we identify the Cepheids adopted in Leavitt's study, compare their pulsation periods with recent catalogs, and verify their classification and pulsation modes. In Sect. \ref{sect:lightcurves} we construct Cepheid light curves from the data collected in Leavitt's notebook \citep{Leavitt_Notebook_1905} and compare them with modern light curves obtained at optical wavelengths. Then, in Sect. \ref{sect:PL} we use recent and well-covered Cepheid light curves to reproduce an updated version of Leavitt's law in the reddening-free $m_{VI}^W$ index. In particular, we describe the effect of the Cepheid sampling in the SMC and the choice of reddening law. Additionally, we discuss the P-L slope obtained by Leavitt and find a bias
brightward at the short period end, due to the non-linearity of the plates and environmental crowding. Finally, we discuss the remarkable accuracy and precision of Leavitt's work in Sect. \ref{sect:conclusion}.  Throughout this paper, we use both Cepheid Period-Luminosity Relation (P-L Relation) and Leavitt Law in order to avoid confusion. \\

\begin{table*}[t!]
%\footnotesize
\begin{center}
\caption{Sample of 25 SMC Cepheids adopted by \citet{Leavitt1912}. \\}
\begin{tabular}{c c c c c c c c}
\hline
\hline
HV ID & OGLE ID & Gaia DR3 ID & RA & DEC & $P^{\rm \, (a)}$  & $P^{\rm \, (b)}$ & $m_{VI}^{W} \, ^{\rm \, (c)}$ \\
    &    & &    (deg) & (deg) & (days) & (days) & (mag) \\
\hline
  HV 1505 & OGLE SMC-CEP-1581 & 4689056005984060032 & 12.7421 & -72.0740 & 1.253 & 1.251 & $15.960$   \\   
 HV 1436 & OGLE SMC-CEP-1113 & 4689029862552916736 & 12.0321 & -72.4331 & 1.664 & 1.656 & $15.784$   \\   
 HV 1446 & OGLE SMC-CEP-1180 & 4689029931272382848 & 12.1376 & -72.4194 & 1.762 & 1.872 & $15.679$   \\   
 HV 1506 & OGLE SMC-CEP-1562 & 4689037898398780544 & 12.7270 & -72.3104 & 1.875 & 1.886 & $15.634$   \\   
 HV 1413 & OGLE SMC-CEP-0770 & 4685829218556335616 & 11.4177 & -73.5713 & 2.174 & 2.154 & $15.473$   \\   
 HV 1460 & OGLE SMC-CEP-1298 & 4689246260148791680 & 12.3241 & -72.0299 & 2.913 & 2.913 & $14.689$   \\   
 HV 1422 & OGLE SMC-CEP-0811 & 4685780668235969152 & 11.5128 & -73.6807 & 3.501 & 3.501 & $14.930$   \\   
 HV 842 & OGLE SMC-CEP-2837 & 4688997358257544192 & 14.6204 & -72.4444 & 4.289 & 4.289 & $14.506$   \\   
 HV 1425 & OGLE SMC-CEP-0915 & 4689027457371425664 & 11.7102 & -72.5433 & 4.547 & 4.547 & $14.377$   \\   
 HV 1742 & OGLE SMC-CEP-2722 & 4685992564797958912 & 14.4179 & -72.5456 & 4.987 & 4.987 & $14.081$   \\   
 HV 1646 & OGLE SMC-CEP-2375 & 4685985795927276928 & 13.8376 & -72.6751 & 5.311 & 5.234 & $14.038$   \\   
 HV 1649 & OGLE SMC-CEP-2384 & 4685985521049473408 & 13.8447 & -72.7043 & 5.323 & 5.324 & $14.124$   \\   
 HV 1492 & OGLE SMC-CEP-1492 & 4689031095213191808 & 12.6072 & -72.4619 & 6.293 & 6.292 & $13.798$   \\   
 HV 1400 & OGLE SMC-CEP-0755 & 4689043361630672768 & 11.3614 & -72.4531 & 6.650 & 6.648 & $13.760$   \\   
 HV 1355 & OGLE SMC-CEP-0404 & 4688847133137909120 & 10.3501 & -73.3613 & 7.483 & 7.482 & $13.613$   \\   
 HV 1374 & OGLE SMC-CEP-0512 & 4685827638001043712 & 10.7716 & -73.5666 & 8.397 & 8.396 & $13.502$   \\   
 HV 818 & OGLE SMC-CEP-0351 & 4685813619225612288 & 10.0985 & -73.6736 & 10.336 & 10.333 & $13.291$   \\   
 HV 1610 & OGLE SMC-CEP-2230 & 4689007047705199872 & 13.6200 & -72.4109 & 11.645 & 11.643 & $13.052$   \\   
 HV 1365 & OGLE SMC-CEP-0423 & 4685821418888717696 & 10.4545 & -73.7284 & 12.417 & 12.409 & $13.246$   \\   
 HV 1351 & OGLE SMC-CEP-0387 & 4685841725497006976 & 10.2661 & -73.5277 & 13.080 & 13.091 & $12.995$   \\   
 HV 827 & OGLE SMC-CEP-1377 & 4689030549749347456 & 12.4283 & -72.4959 & 13.470 & 13.464 & $12.788$   \\   
 HV 822 & OGLE SMC-CEP-0431 & 4685838907998450944 & 10.4812 & -73.5399 & 16.750 & 16.742 & $12.544$   \\   
 HV 823 & OGLE SMC-CEP-0574 & 4685823721000358144 & 10.9519 & -73.6135 & 31.940 & 31.928 & $11.365$   \\   
 HV 824 & OGLE SMC-CEP-0921 & 4688974886982934016 & 11.7213 & -72.7144 & 65.800 & 65.937 & $10.111^{\rm \, (d)}$   \\   
 HV 821 & OGLE SMC-CEP-0417 & 4685821487608191104 & 10.4310 & -73.7233 & 127.000 & 128.197 & $9.315^{\rm \, (d)}$   \\  
 \hline
\end{tabular}
\label{table:names_periods}
\end{center}
{\textbf{Notes:} (a) Original pulsation period from \citet{Leavitt1912}. (b) Pulsation period from the OGLE survey \citep{Soszynski2015}. (c) Apparent mean magnitude in the optical Wesenheit index, $m_{VI}^{W} = I-1.278 \, (V-I)$, built with $V$ and $I$ magnitudes from OGLE \citep{Soszynski2015} assuming $R_V = 2.74$ (see Sect. \ref{subs_3_2}). (d) These $V$-band mean magnitudes are not available in OGLE so they are taken from \citet{Henden2015}. \\ }
~ \\
\end{table*}

\newpage
\section{Cepheid sample} 
\label{sect:identification}

%\subsection{Identifying the SMC Cepheids studied by Henrietta Leavitt} 

The first step of this analysis is to identify the 25 SMC Cepheids used to derive the original P-L relation from \cite{Leavitt1912} (see Fig.~\ref{fig:PL_Leavitt}). The Cepheid names listed in her paper are in the Harvard Variable (HV) system. Their corresponding names in the OGLE survey \citep{Soszynski2015} and in the \textit{Gaia} DR3 catalog can be found using the Simbad\footnote{\href{https://simbad.u-strasbg.fr/simbad/}{https://simbad.u-strasbg.fr/simbad/}} database. Their pulsation periods are provided in the OGLE survey and in the \textit{Gaia} DR3 catalog \citep{GaiaDR3contents}, they are listed in Table~\ref{table:names_periods}.

For 16 of Leavitt's SMC Cepheids, Table VI of \cite{Leavitt1907} gives information about the quality of the time-series observations. They were observed between 1889 and 1906 at an average of $\sim 41$ times each. All had at least 20 observations, and two had more than 80. While these are several times the number of observations that are typically used to obtain periods for Cepheids for modern cosmic distance ladders \citep{Freedman2001, Riess2022}, it is important to remember that over such a long baseline, the observational sampling was likely to be quite sparse, \new{which can increase the light curve scatter due to potential small period changes}. In extreme cases, such as that of HV 1436, its 22 epochs of observations covered more than 2,800 pulsational cycles. \new{Modern observations usually sample several phases of the same pulsation cycle for a variable star, which avoids aliasing and helps narrow down the period measurement.}

\begin{figure}[t!]
\centering
\includegraphics[height=5.6cm]{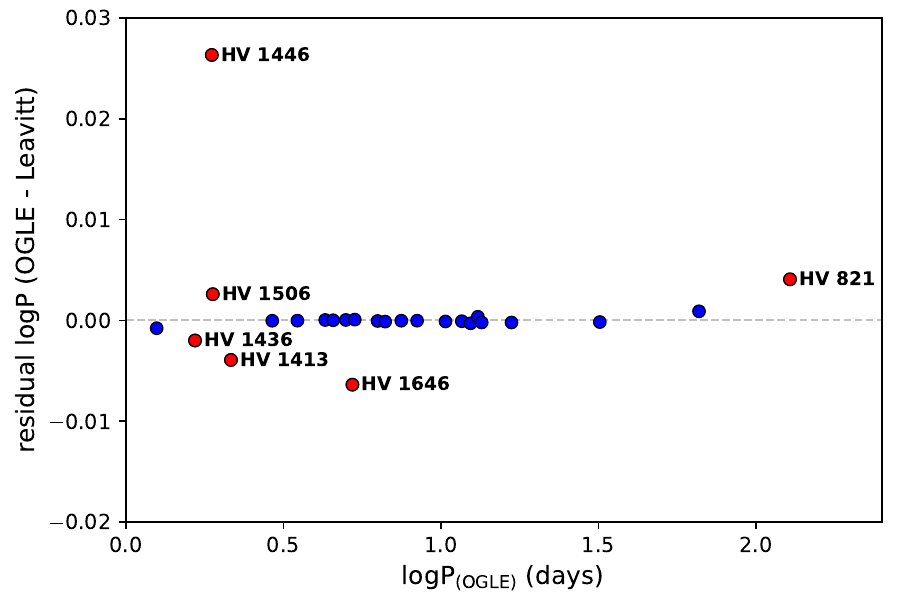} 
\caption{Residuals between the periods (expressed in $\log P$) from \citet{Leavitt1912} and from OGLE IV \citep{Soszynski2015}.}
\label{fig:residual_periods}
\end{figure}

Figure~\ref{fig:residual_periods} shows a comparison between the Cepheid pulsation periods measured by \citet{Leavitt1912} from photographic plates, and those from the OGLE catalog \citep{Soszynski2015}. Surprisingly, given the sparsity of Leavitt's original data, the agreement is very good for most Cepheids. The largest outlier in $\log P$ is HV 1446 (red point in the top left corner of Fig. \ref{fig:residual_periods}), whose 21 observations spanned 2,052 cycles. Even for this variable star, the measured periods were still within $\sim 0.1$ days: Leavitt found a period of 1.76 days ($\log P = 0.246$) while OGLE reports 1.87 days ($\log P = 0.272$). Except for the 6 Cepheids indicated in red and labeled in Fig. \ref{fig:residual_periods}, the agreement is excellent (to 0.01 day on average). In Appendix \ref{appendixA}, we discuss the particular case of BZ Tuc (HV 821) and its various period measurements and potential evolution. 

%\begin{figure}[t!]
%\centering
%\includegraphics[height=6.0cm]{OC_BZ_Tuc.pdf} 
%\caption{O-C diagram of HV 821 (BZ Tuc) constructed using ASAS time series collected over a baseline of 10 years. O-C is expressed as a fraction of the fit period. Uncertainties are estimated as $\sim \sigma_{phs}/\sqrt{N_{obs}}$ where $\sigma_{phs}$ is the phase uncertainty $N_{obs}$ is the number of observations per cycle. Only cycles observed near peak are included.}
%\label{fig:OC_BZ_Tuc}
%\end{figure}

%This would correspond to a period change of $895 \pm 84$ s/yr, measured over a baseline of 115 years corresponding to about 330 pulsation cycles.  \\

% \footnote{\uline{\href{http://www.astrouw.edu.pl/asas/?page=aasc}{http://www.astrouw.edu.pl/asas/?page=aasc}}}

% \subsection{Pulsation modes and classifications} 

The paper by \citet{Leavitt1912} does not provide any classification or pulsation mode for their 25 SMC Cepheids. In fact, at the time, Cepheids were not yet classified between different types and were considered as a unique population. For this reason, and because he was using a mix of classical and Type II variables, Edwin Hubble's first estimate of the distance to galaxies was overestimated and his initial value for the Hubble constant was around 500 km/s/Mpc. According to the OGLE catalog \citep{Soszynski2015}, all Cepheids from \cite{Leavitt1912} are fundamental mode pulsators, although some of them have very short periods (three Cepheids with $\log P < 0.25$) and could be suspected to be first overtone pulsators or to be incorrectly classified as classical Cepheids. Additionally, all 25 SMC Cepheids, even the ones with very short periods, are confirmed fundamental-mode classical Cepheids by the Gaia DR3 reclassification \citep{Ripepi2022Gaia}. \\

\begin{figure}[t!]
\centering
\includegraphics[width=8.5cm]{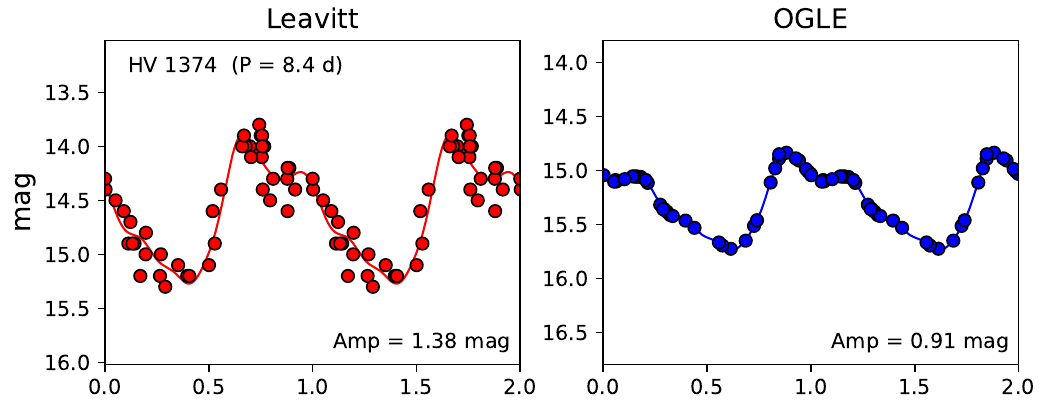} \\
\vspace{0.3cm}
\includegraphics[width=8.5cm]{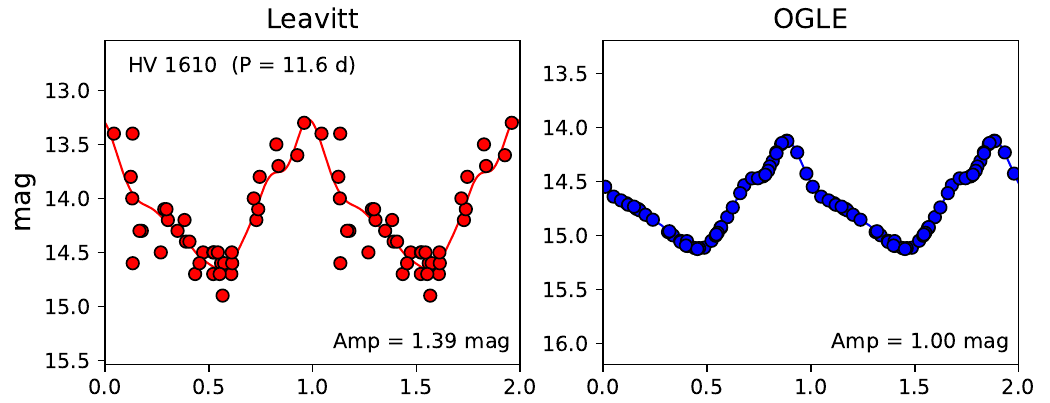}  \\
\vspace{0.3cm}
\includegraphics[width=8.5cm]{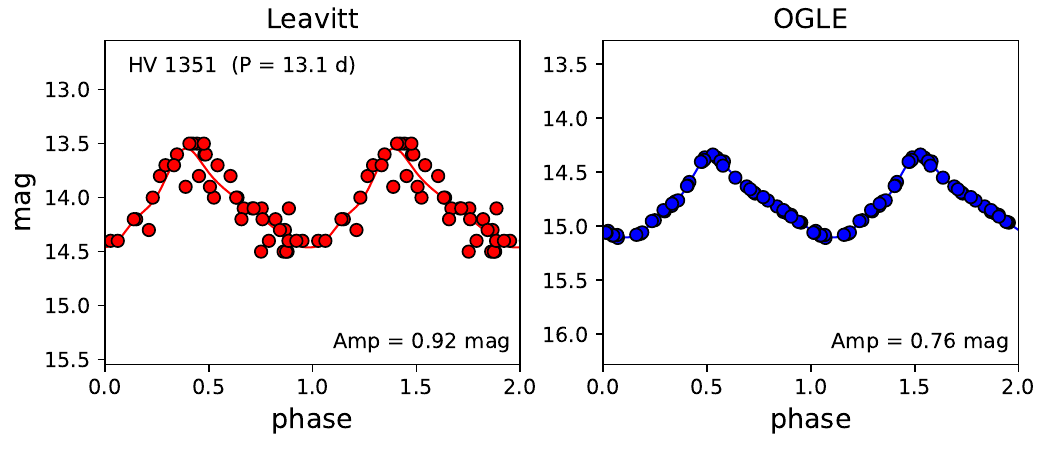} 
\caption{Example of Cepheid light curves extracted from Henrietta Leavitt's notebook \citep{Leavitt_Notebook_1905} in red and from the OGLE IV catalog \citep{Soszynski2015} in blue. The light curve shape from OGLE was fitted to Leavitt's data points.}
\label{fig:leavitt_light_curves}
\end{figure}

\section{Cepheid light curves}
\label{sect:lightcurves}

At the time of these observations, a standard scale of magnitudes had not yet been developed \citep{Pickering_1909}. Thus, all of the measurements in Leavitt's notebooks and 1912 paper were recorded in \textit{provisional} magnitudes. However, Leavitt hypothesized that converting the provisional magnitudes of Cepheids to a standard scale could further decrease the scatter in the P-L relation, writing, \textit{``It is possible that the deviations from a straight line may become smaller when an absolute scale of magnitudes is used''} \citep{Leavitt1912}. 

Leavitt's published papers \citep{Leavitt1907, Leavitt1912} do not discuss how her Cepheid light curves were constructed. However, her notebooks \citep{Leavitt_Notebook_1905} contain a handwritten collection of Cepheid measurements in the same regions of the SMC with dozens of photographic plates obtained at Harvard's Boyden Station located in Arequipa, Peru. Using the measurements reported from pages 4 to 40 of her 1905 notebook (\new{see Appendix \ref{appendixB}}), we were able to extract and reconstruct the majority of the original observations for all 25 Cepheids of her sample. These light curves are based on 34 epochs on average, with a minimum and a maximum of 17 and 56 epochs, respectively. We note that additional observations listed in other notebooks might be available but were not included here. These original light curves were then compared with modern and well-sampled $V$-band light curves from OGLE \citep{Soszynski2015}. For 17 of the 25 Cepheids, we were able to determine a clear periodicity and resemblance to light curve shapes of these stars from OGLE. Three Cepheids are displayed in Fig.~\ref{fig:leavitt_light_curves}  as an example. They are representative of the 17 conclusive recovered light curves. The similarity between both data sets is remarkable and even clearly shows the Hertzsprung progression, a small local maximum  whose phase depends on the period of the Cepheid \citep{Hertzsprung1926}.

\begin{figure}[t!]
\centering
\includegraphics[height=5.1cm]{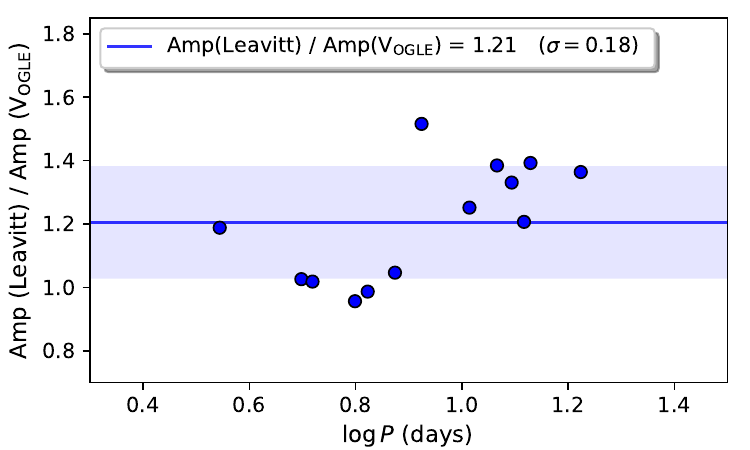} 
\caption{Cepheid light curve amplitude ratios between Leavitt's provisional scale and the OGLE $V$-filter. }
\label{fig:amp_ratios}
\end{figure}

While the exact wavelength of Leavitt's measurements is unknown, it is possible to infer the approximate wavelength by examining the amplitude of light curves. Cepheid amplitudes are known to correlate with wavelength, with larger amplitudes in the blue and smaller amplitudes in the infrared. Today, amplitude ratios between different filters have been established with great precision \citep{Klagyivik2009, Riess2020, Sharon2024}. Thus, we use the amplitude ratios between Leavitt's system and OGLE's $V$-band light curves to derive the wavelength of her observations. We first fit the well-sampled OGLE light curves with Fourier series. Then, we adopt OGLE's light curve shapes (which are similar at similar wavelengths) and set the amplitude as a free parameter to fit Leavitt's light curves. We compare amplitudes in both systems and we find that amplitudes in Leavitt's provisional scale of magnitudes are 1.21 times larger than in the $V$-band OGLE filter (see Fig. \ref{fig:amp_ratios}). Only Cepheids with periods in $0.5 < \log P < 1.5$ were considered in this analysis, in order to exclude faintest and brightest variables which might be affected by the non-linearity of the photographic plates response and possibly crowding. \new{A possible dependence of the amplitude ratio with period is discussed in Sect. \ref{subs_3_1}.} The fact that the amplitude ratio between Leavitt’s filter and the $V$-band was greater than 1 suggests that the effective wavelength of her observations was shorter than the modern $V$ band. We can further narrow this range down by comparing with the detailed amplitude ratios analysis from \citet{Klagyivik2009} which derived that $A(B) \sim 1.5 ~A(V)$ with a small dependence on the period. Therefore we conclude that Leavitt's provisional scale of magnitude is comparable to a blue filter between the $B$ and $V$ bands. An additional discussion on the wavelength of Leavitt's observations is presented in Sect. \ref{subs_3_1}. \\

\section{Period-Luminosity relation} 
\label{sect:PL}

In Sect.~\ref{subs_3_1}, we reproduce the original P-L relation from \citet{Leavitt1912} by simply using their original magnitudes, but we adopt the pulsation periods from the OGLE survey \new{which are more reliable and based on better data}. We compare the P-L slope to what we expect it to be at the $B$-band wavelength and attribute differences to the non-linearity of the photographic plate response and to crowding. In Sect.~\ref{subs_3_2}, we update Leavitt's photometry with the most recent light curves from the OGLE survey and we describe the impact of the spatial sampling of SMC Cepheids.    \\

\begin{figure*}[t!]
\centering
\includegraphics[height=14.5cm]{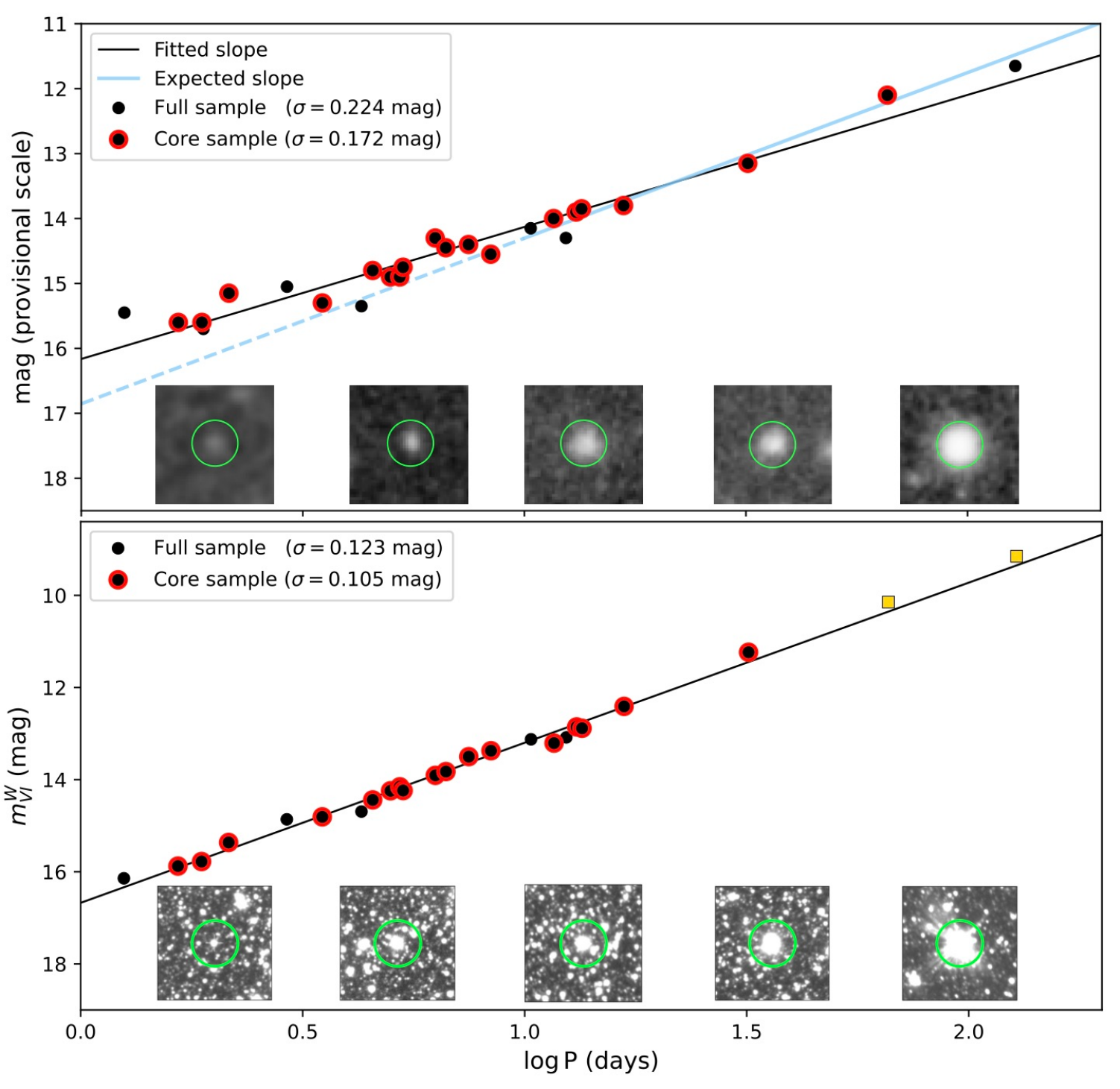} 
\caption{{\bf Top}: P-L relation obtained in a \textit{provisional scale of magnitudes} using the mean between the minimum and maximum from Leavitt's original data \citep{Leavitt1912}. The solid black line is the P-L relation fit to the subset of Cepheids within 0.8 deg of the SMC center (red circles). The solid blue line shows the P-L relation fit to Cepheids with $\log P > 1$, with slope fixed to the modern $B$-band value of -2.55 (Fig. \ref{fig:PL_slopes}). The dashed-blue line shows the extension of this fit to $\log P < 1$. The stamps show five Cepheids of different brightness in this sample, obtained from the digitized photographic plates by the \textit{DASCH} project. {\bf Bottom}: PL relation for the same Cepheid sample as above, except that mean magnitudes are expressed in the $m_{VI}^W$ Wesenheit index (a combination of $V$ and $I$ filters) and mean magnitudes taken from the OGLE survey. The two longest period Cepheids are shown in orange and are not considered in the fit, as their $V$-band magnitudes are not available in the OGLE survey. We assumed a reddening parameter of $R_V=2.74$ and included the geometry correction from \citet{Graczyk2020}. The stamps are taken from the OGLE survey and were obtained with the 1.3-m Warsaw telescope at Las Campanas Observatory in Chile. \\}
%{\bf Top}: P-L relation obtained in a \textit{provisional scale of magnitudes} using the mean between the minimum and maximum from Leavitt's original data \citep{Leavitt1912}. Using only Cepheids within 0.8 deg from the SMC center (red circles) reduces the P-L scatter from 0.22 mag to 0.17 mag. The solid black line is the fitted P-L relation. The blue line shows the expected slope assuming Leavitt's observations were equivalent to the $B$ band's wavelength. Only Cepheids with $\log P>1$ were considered in this fit. At short periods, Cepheids appear brighter than expected due to the nonlinearity of photographic plates. The stamps show five Cepheids of different brightness obtained from the digitized photographic plates by the DASCH project. {\bf Bottom}: PL relation for the same Cepheid sample as above, except that mean magnitudes are expressed in the $m_{VI}^W$ Wesenheit index (a combination of $V$ and $I$ filters) and taken from the OGLE survey. We assumed a reddening parameter of $R_V=2.74$ and included the geometry correction from \citet{Graczyk2020}. Using only Cepheids within 0.8 deg from the SMC center (red circles) reduces the P-L scatter from 0.13 mag to 0.11 mag. The stamps are taken from the OGLE survey and were obtained with the 1.3-m Warsaw telescope at Las Campanas Observatory in Chile.  \\}
\label{fig:two_PL_relations}
\end{figure*}

\subsection{P-L relation with magnitudes from Leavitt \& Pickering (1912)}
\label{subs_3_1}

\citet{Leavitt1912} only provide the minimum and maximum observed magnitudes for the 25 Cepheids in their Table 1 (see Fig. \ref{fig:PL_Leavitt}, left) but do not mention using the full light curves. Therefore, with only the information available in \citet{Leavitt1912}, we adopt the mean of the minimum and maximum magnitude for each Cepheid and reconstruct their P-L relation in the top panel of Fig.~\ref{fig:two_PL_relations}. With these adjustments, the Leavitt law obtained with her original data has a scatter of $0.22 \, \rm mag$. As a comparison, in the modern {\it Hubble} Space Telescope (HST) filters, \cite{Breuval2024} find the lowest dispersion to date for the SMC Leavitt law with a scatter of 0.10 mag based on 87 Cepheids in the core of the SMC.

We also investigate the impact of the spatial distribution of Leavitt's Cepheid sample in the SMC. \citet{Breuval2022, Breuval2024} show that the elongated shape of the SMC produces additional scatter in the Leavitt law. For this reason, it is recommended to limit the Cepheid sample to a narrow region around the SMC center. Progressively reducing the sample to $R = 1 ^{\circ}$, $0.9 ^{\circ}$ and $0.8 ^{\circ}$ decreases the scatter of the Leavitt law to $0.20 \, \rm mag$ (N = 23), $0.18 \, \rm mag$ (N = 22), and $0.17 \, \rm mag$ (N = 18) respectively. Limiting the sample to an even narrower region would discard more than half of Leavitt's sample, therefore, in the following we assume that the ideal subsample for this study would be a region of radius $0.8^{\circ}$ around the SMC center and use this subset (red circles) for the P-L relation fit (shown in black) in Figure \ref{fig:two_PL_relations}.  This test confirms that Cepheids located outside of the SMC core increase the P-L scatter \new{(see also Sect. \ref{subs_3_2})}, and shows that after excluding the latter, Leavitt's P-L scatter is comparable to that obtained with modern data. The P-L scatter for different regions around the SMC center is listed in the upper part of Table \ref{table:improvements}.

\begin{figure}[t!]
\centering
\includegraphics[height=6.6cm]{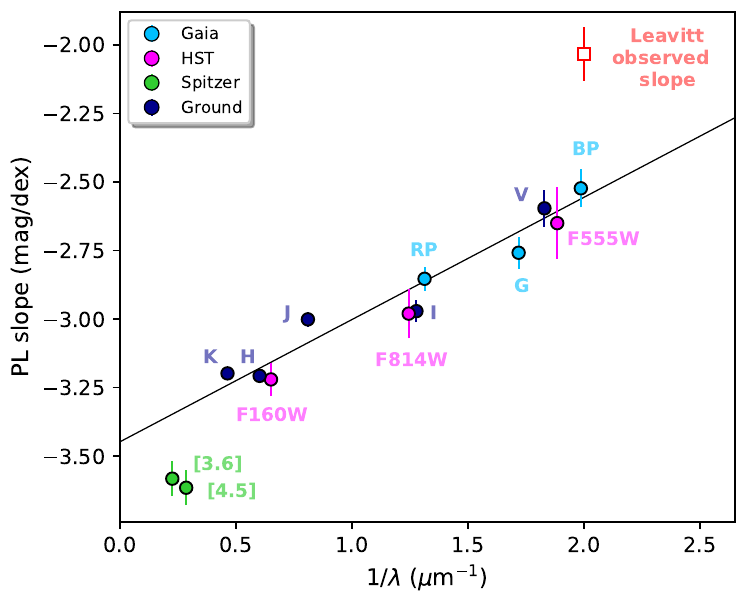} 
\caption{P-L slope for the SMC relations as a function of wavelength, labeled with the filters. Values in \textit{Gaia}, \textit{Spitzer} and ground-based filters are from \citet{Breuval2022}, and values in HST filters are from \citet{Breuval2024}. The red square shows the observed P-L slope in Leavitt's provisional scale of magnitude ($-2.033 \pm 0.098$ mag/dex) at the position of the expected wavelength of photographic plate observations (0.5 $\mu$m).  \\  }
\label{fig:PL_slopes}
\end{figure}

The color response of Harvard photographic plates, including those used by Leavitt for her discovery, is comparable to the Johnson $B$ filter \citep{Tang_2013}. This can be verified empirically by comparing Leavitt's measurements with modern data. For example, in Sect.~\ref{sect:lightcurves}, the amplitude ratios between Cepheid light curves from Leavitt's notebook and from the OGLE survey point towards a wavelength between the $B$-band and the $V$-band, around $\lambda \sim 0.50 \, \mu$m. In principle and assuming accurate magnitude measurements over a wide dynamic range of 5 magnitudes, we might infer the wavelength of the photographic plates by measuring the P-L slope, which is strongly correlated with the wavelength at which it is measured, with steeper slopes in the infrared and shallower slopes in the optical. For a plate response near the $B$-band, Leavitt's P-L slope is expected around $-2.55$ mag/dex (see Fig.~\ref{fig:PL_slopes}). In practice, the observed slope of Leavitt's original P-L relation (Fig.~\ref{fig:two_PL_relations}, upper panel, solid black line) is much shallower, with $-2.03 \pm 0.11$ mag/dex. It is very likely that the shallow P-L slope obtained by Leavitt is a consequence of two systematic errors, 1) the well-known non-linearity of the photographic plates over a wide dynamic range and 2) environmental crowding, exacerbated by low resolution.

\begin{table}[t!]
%\small
%\centering
\caption{Successive improvements in the scatter of the P-L relation from Leavitt's Cepheid sample. The columns are: (1) the radius $R$ of the region around the SMC center, (2) the number $N$ of Cepheids, (3) the inclusion or not of geometry corrections, (4) the value of $R_V$ and (5) the P-L scatter $\sigma$ in mag. \\  }
\begin{tabular}{c c c c c }
\hline
\hline
$R$     & $N$ & geo. corr. & $R_V$ & $\sigma$    \\
(deg) &   &            &       & (mag)      \\
\hline
\multicolumn{5}{c}{Leavitt's provisional magnitude scale} \\
\hline
Full              & 25 & no & -- & {\bf 0.224}  \\
$R < 1.0^{\circ}$ & 23 & no & -- & {\bf 0.197}  \\
$R < 0.9^{\circ}$ & 22 & no & -- & {\bf 0.179}  \\
$R < 0.8^{\circ}$ & 18 & no & -- & {\bf 0.172}  \\
\hline
\multicolumn{5}{c}{OGLE mean magnitudes ($W_{VI}$)} \\
\hline
Full              & 23 & no  & 3.3 & {\bf 0.137}  \\
Full              & 23 & yes & 3.3 & {\bf 0.126}  \\
$R < 1.0^{\circ}$ & 21 & yes & 3.3 & {\bf 0.118}  \\
$R < 0.8^{\circ}$ & 17 & yes & 3.3 & {\bf 0.112}  \\
$R < 0.8^{\circ}$ & 17 & yes & 2.74 & {\bf 0.105}  \\
\hline
\end{tabular}
\label{table:improvements}
\end{table}

\begin{table*}[t!]
%\small
%\centering
\caption{Leavitt law ($m_{VI}^W = \alpha \log P + \beta$) obtained with Leavitt's Cepheid sample for different $R_V$ values, with periods and magnitudes taken from the OGLE catalog, after excluding the two longest period Cepheids which do not have $V$-band photometry in OGLE.}
\begin{tabular}{c c c | c c c | c c c | c c c c}
~ \\
\hline
\hline
$\alpha$ & $\beta$ & $\sigma$ & $\alpha$ & $\beta$ & $\sigma$ & $\alpha$ & $\beta$ & $\sigma$ & sample & $N$   \\
\hline
\multicolumn{3}{c}{$R_V = 3.3$} & \multicolumn{3}{c}{$R_V = 3.1$} & \multicolumn{3}{c}{$R_V = 2.74$} & \\
\hline
$-3.459$ & 16.548 & 0.126 & -3.441 & 16.575 & 0.125 & -3.407 & 16.626 & 0.123 & full              & 23 \\
$-3.526$ & 16.617 & 0.118 & -3.509 & 16.645 & 0.116 & -3.478 & 16.698 & 0.114 & $R < 1.0^{\circ}$ & 21 \\
$-3.529$ & 16.697 & 0.112 & -3.510 & 16.624 & 0.109 & -3.475 & 16.674 & 0.105 & $R < 0.8^{\circ}$ & 17 \\
\hline
\end{tabular}
{\flushleft \textbf{Notes:} In the modern literature, the P-L slope in the SMC and in the $m_{VI}^W$ index is found between $-3.32$ \citep{Breuval2022} and $-3.46$ mag/dex \citep{Soszynski2015}. The steeper slope found by \cite{Soszynski2015} is mostly due to their use of short period Cepheids and a different coefficient for the Wesenheit index.  }
\vspace{0.4cm}
\label{table:PL}
\end{table*}

The sensitivity of a photographic plate is described by a \textit{characteristic curve} (or Hurter-Driffield curve), which is commonly described as function of $\log$(Exposure) versus $\log$(Density). Modern texts often represent Exposure as $H$, and it is defined as,
\begin{equation}
H = t \times I
\end{equation}
where $t$ is the exposure time and $I$ is the intensity of the incident light. Density ($D$), on the other hand, describes the optical density of the photographic plate as a result of the development process. In $\log-\log$ space, the characteristic curve generally has an ``S" shape. In the ``toe region" where there is low exposure, the photographic plate response is nonlinear and relatively flat. The mid-exposure region has a predictable, linear relationship between $D$ and $H$. The high exposure region is known as the shoulder -- here saturation causes the characteristic curve to be flat again. The procedures and emulsions used will affect the shape of the characteristic curve. As explained in \cite{Laycock_2010}, the photographic plate collection is a heterogeneous dataset and there is scant information about the procedures, emulsions, or developers. Thus, the characteristic curve is likely to be different for each plate, though we can assume that the ``S" shape will be present in the sensitivity. It is beyond the scope of this work to derive a new photometric calibration for each plate used here, and also orthogonal to our main goal of specifically re-examining Leavitt's own work. Empirically, these nonlinearities can have the effect of compressing the observed magnitude range, e.g., from a true $\Delta m$ from faintest to brightest of $\sim$5 to a measured $\Delta m \sim$ 4 \citep{Laycock_2010, Tang_2013}. This compressed magnitude range over the same period range could be responsible for the relatively flatter slope in Leavitt's results compared to modern measurements. However, if we limit the magnitude range of the Cepheids to the subset that fall within the linear regime of the photographic plate, we expect that we will see better agreement in the measurement of the slope, even in Leavitt's provisional scale of magnitude. To test this hypothesis, we excluded faint and bright Cepheids outside of the range $0.5 < \log P$ (days) $< 1.5$ which are most likely to be affected by this non-linearity: the resulting slope measured in Leavitt's P-L relation becomes steeper, $-2.19 \pm 0.11$ mag/dex, and in better agreement with the expected value.

This ``toe" and ``shoulder" effects can potentially also be seen in the amplitude ratios shown in Figure \ref{fig:amp_ratios}. For the faintest Cepheids with $\log P \lesssim 0.9$, the amplitude ratios all fall below the mean of 1.2. Greater than $\log P \gtrsim 0.9$, the amplitude ratios are all larger than the mean. It is possible that the shortest period Cepheids are either not detected at the minima in their pulsational phases, or fall in the regime of the toe effect at their minima, which would lead to a compressed amplitude. We illustrate this potential effect in the top panel of Figure \ref{fig:two_PL_relations}: we fit the P-L relation (solid blue line) with the expected slope based on modern observations to all of the Cepheids with $\log P > 1$. Even using only Leavitt's provisional scale of magnitudes, points qualitatively appear to fit the modern slope quite well. When we extend the relation to $\log P < 1$ (dashed blue line) we see that the shorter period end appears to be flattened compared to the modern version. Together, these observations suggest that at least some of the differences in the steepness of the slope compared to modern measurements can be attributed to the nonlinearity of the photographic plate response.

In addition to the non-linear response of photographic plates, the possibility of a higher level of crowding for fainter Cepheids could also explain the shallow slope observed by Leavitt. Indeed, short period Cepheids are barely visible on photographic plates and the presence of nearby stars around the Cepheids is very clear on Fig.~\ref{fig:two_PL_relations}. To test this hypothesis, we performed aperture photometry on OGLE images (A. Udalski, private communication) of Leavitt's Cepheids using different aperture sizes. We find that contamination by nearby stars is about five times larger at $\log P < 0.25$ than at $\log P > 1.30$. These short-period Cepheids were likely measured brighter than they actually are, resulting in a shallower slope at the faint end of the P-L relation.  \\

\begin{figure*}[t!]
\centering
\includegraphics[height=8.9cm]{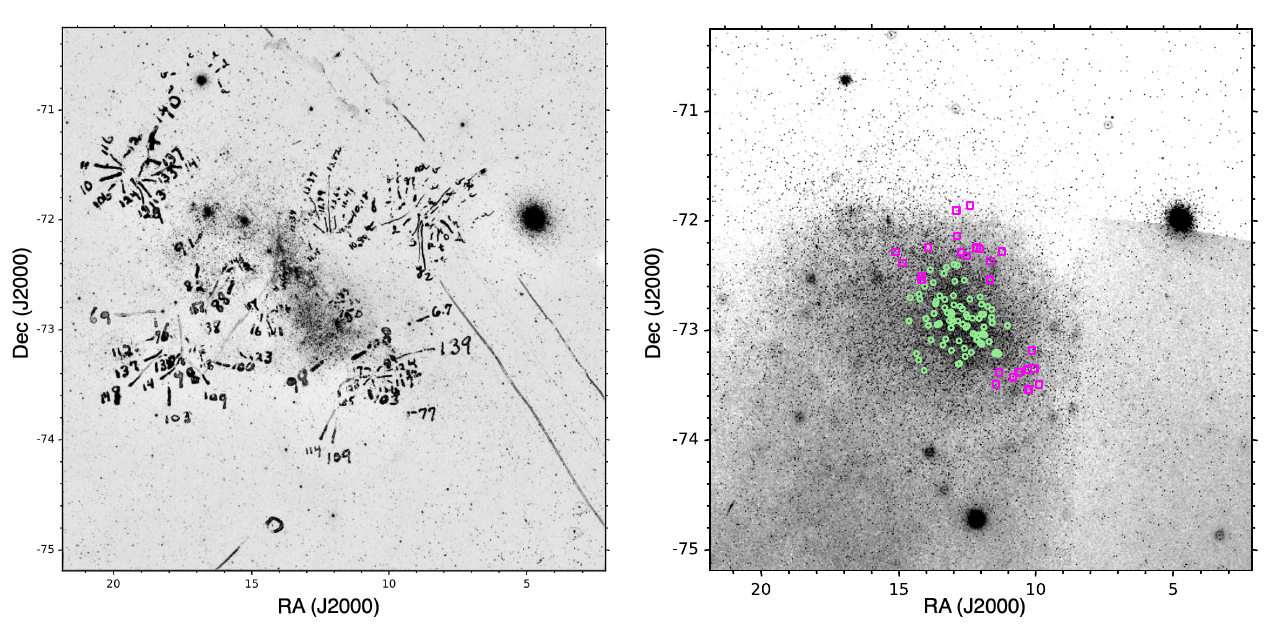} 
\caption{\textbf{Left}: Photographic plate image provided by the Harvard Plate Stacks, Center for Astrophysics, Harvard \& Smithsonian. This plate (B20678) of the Small Magellanic Cloud was analyzed by Leavitt for her study of variable stars, which led to her discovery \citep{Leavitt1912}. \textbf{Right}: Same region as on the left, but from DSS2-IR. The light green circles are the SMC Cepheids that were observed by HST/WFC3 in \cite{Breuval2024}. The magenta squares are the 25 Cepheids from Leavitt's study. The orientation and scale of the two maps are identical. \\}
\label{fig:SMC_map}
\end{figure*}

\subsection{P-L relation with magnitudes from OGLE}
\label{subs_3_2}

In this section, we adopt the $V$ and $I$ mean magnitudes from the OGLE IV survey \citep{Soszynski2015} for the same sample of 25 SMC Cepheids used by Leavitt. For the two longest period Cepheids, the $V$-band mean magnitudes are taken from \citet{Henden2015} as they are not available in OGLE IV, they were not considered in the P-L fit. To mitigate the effect of dust\footnote{~$E(B-V) \sim 0.05$ mag in the SMC \citep{Skowron2021}.}, we derive a reddening-free Wesenheit index defined as $m_{VI}^W = I - R \, (V-I)$. Assuming the reddening law from \citet{Fitzpatrick1999} and $R_V = 3.1$ (as usually adopted in the Milky Way) yields $R = 1.387$. However, \citet{Gordon2003} reported a value of $R_V = 2.74$ in the SMC, which gives $R = 1.278$. Finally, the value adopted in the SH0ES distance ladder \citep{Riess2022} is $R_V = 3.3$, which gives $R=1.445$.  The following results are given for all three $R_V$ values in Table \ref{table:PL}.

Similarly to Sect. \ref{subs_3_1}, using only Cepheids within $0.8^{\circ}$ from the SMC center reduces the P-L scatter from 0.13 to 0.11 mag. \citet{Breuval2021, Breuval2024} show that, additionally to excluding Cepheids located far from the SMC core, adopting a geometric model of the SMC to correct for residual depth effects further reduces the scatter of the Leavitt law: we apply the SMC model from \cite{Graczyk2020} so that each Cepheid is considered at its own individual distance depending on its position, contrary to previous studies \citep[e.g.][]{Soszynski2015, Wielgorski2017, Gieren2018} which assumed the same distance for all SMC Cepheids. The Leavitt law obtained for the three values of $R_V$ and for different regions around the SMC center are listed in Table \ref{table:PL}. Adopting $R_V = 2.74$ \citep{Gordon2003} slightly reduces the P-L scatter compared with using $R_V = 3.1$ or $R_V = 3.3$, but only represents a minor improvement (see lower part of Table \ref{table:improvements} for the successive improvements in the P-L scatter). Our final P-L relation based on Leavitt's sample and modern photometry is shown in the bottom panel of Fig. \ref{fig:two_PL_relations}, with periods and magnitudes from OGLE, $R_V = 2.74$ and geometry corrections from \cite{Graczyk2020}'s planar model of the SMC based on eclisping binaries. 

Below each P-L relation shown in Figure \ref{fig:two_PL_relations}, we have also shown cutouts of the field surrounding five of the Cepheids from the SMC sample (HV 1446, HV 1400, HV 1351, HV 823, HV 821) at a range of periods. The images are taken from both the digitized plates from the \textit{DASCH} project \citep{Grindlay_2012} and also from the OGLE survey to give a sense of the crowding in the field around each representative star. The photographic plate cutouts were extracted using the {\tt daschlab} Python package from \textit{DASCH} Data Release 7 \citep{Williams_2025}. The software is available on Github\footnote{\texttt{daschlab} codebase: \url{https://github.com/pkgw/daschlab}} and is archived in Zenodo \citep{williams_2024_14574817}. \\

\section{Discussion}
\label{sect:conclusion}

This paper aims to highlight the quality of the work performed by Henrietta Leavitt in \cite{Leavitt1912} and the significant improvements of our knowledge, both in astronomy in general and on the Cepheid Leavitt law, since her pioneering discovery of the Leavitt law. The decrease by a factor of two in dispersion between the first P-L relation derived by Leavitt and the present reanalysis with OGLE data (Table \ref{table:improvements}) can be explained by the tremendous advancement of photometry over the last century, in particular the linearity of modern detectors compared to photographic plates and the availability of fully covered light curves in standard photometric systems for a large number of Cepheids, as compared with visual inspection of photographic plates performed by Leavitt. \newc{As Leavitt herself noted, ``\textit{The measurement and discussion of these objects present problems of unusual difficulty, on account of the large area covered by the two regions, the extremely crowded distribution of the stars contained in them, the faintness of the variables, and the shortness of their periods}" \citep{Leavitt1912}.}

\newc{In addition to these technological advances, developments in our astrophysical knowledge have also contributed to the reduction in the P-L scatter. These include the relatively recent use of the reddening-free Wesenheit index and an improved understanding of the geometry of the SMC and its depth effects \citep{Breuval2022}.}  % cite some papers about the Wesenheit and the depth effects. 

With the \textit{Hubble} Space Telescope program GO-17097 (PI: A. Riess, Cycle 30), \cite{Breuval2024} provide consistent WFC3 photometry for 88 SMC Cepheids and improve the local measurement of the Hubble constant by including the SMC as the fourth anchor of the SH0ES distance ladder. This sample focuses on long period Cepheids ($P>6$ days) in the inner region of the SMC ($R<0.6^{\circ}$), and unfortunately none of Leavitt's variables are included in this sample (see Fig.~\ref{fig:SMC_map}).

To place the quality of Leavitt's work in a historical context, the first Cepheid Leavitt law can be compared to Hubble's first velocity-distance diagram. Because Hubble's measurements were affected by serious errors due to confusion between different types of Cepheids, modern distances to Hubble's galaxies are seven times larger than his original values. As a result, his first estimate of the expansion rate of the universe, later called the Hubble constant, was about 500 km/s/Mpc, almost seven times larger than currently measured. Nevertheless, all his distances were underestimated by about the same factor, so his conclusion about the expanding universe was still valid. On the other hand, Leavitt's first measurements of Cepheid light curves and periods were already extremely accurate and agree very well with modern data. The 25 Cepheids used by Leavitt are still listed in widely used catalogs such as OGLE and \textit{Gaia}, and are all classified as fundamental mode pulsators.

Leavitt’s discovery had a tremendous impact on our understanding of the universe. By allowing astronomers to determine distances to objects too distant for parallax, the Leavitt law opened the path to the discovery of the extragalactic universe in the 1920s, and subsequently, the expansion of the universe and Hubble's law. The latest measurements of the Hubble Constant using Cepheids have achieved a 1.2\% precision \citep{Breuval2024, Riess2022} and play an integral role in characterizing the current Hubble tension. A hundred and thirteen years after the publication of her paper, and thanks to her major breakthrough, Cepheid distance measurements and Leavitt’s law remain at the forefront of observational cosmology. \\

% \begin{figure*}[h!]
% \centering
% \includegraphics[width=0.49\textwidth]{HV_1400_1912.pdf} \includegraphics[width=0.49\textwidth]{HV_1400_1912.pdf} 
% \caption{{\it Left:} Placeholder for example 1912 figure from Leavitt's paper {\it Right:} Placeholder for example figure from modern analysis. Check the HV catalog number for sources }
% \label{fig:HV827}
% \end{figure*}

\section*{Acknowledgements}

We would like to thank the members of Project PHaEDRA, Wolbach Library, and the Smithsonian Institution Transcription Center for helpful discussions and providing historical context and transcriptions of Leavitt's notebooks. In particular, we thank Katie Frey, Riley Rhiannon, Giancarlo Romeo, Emily R. Cain, and the Smithsonian Institution Transcription Center volunteers that transcribed the digitized notebooks. We would also like to thank Peter K. G. Williams of the \textit{DASCH} Project for providing the digitized photographic plates of the SMC and for his helpful discussions regarding the nonlinearities in photographic plate magnitudes and in extracting the images from the photographic plates. This material is based on work supported by the National Science Foundation Astronomy \& Astrophysics Postdoctoral Fellowship under Grant No. 2401770. We are grateful to Igor Soszy\'nski and Andrzej Udalski for providing a sample of images from the OGLE survey for Leavitt's Cepheid sample. This work has made use of data provided by Digital Access to a Sky Century @ Harvard (\textit{DASCH}), which has been partially supported by NSF grants AST-0407380, AST-0909073, and AST-1313370. Work on \textit{DASCH} Data Release 7 received support from the Smithsonian American Women’s History Initiative Pool. \\

\bibliography{Breuval_2025.bib}{}
\bibliographystyle{aasjournal}

%%% 

\appendix
\section{Note On the Periodicity of BZ Tuc}
\label{appendixA}

One variable of particular interest in Leavitt's sample is BZ Tuc (HV 821), which showed one of the largest discrepancies in period when comparing the result obtained by modern authors and Leavitt. This is also the Cepheid with the longest period in Leavitt's 1912 work. Depending on the catalog, BZ Tuc has a period of 127 days \citep{Leavitt1912}, 128.13428 days \citep{GaiaDR3contents} or 128.197 days \citep{Soszynski2015}, respectively $\log P =$ 2.1038, 2.1077, and 2.1079. Compared to other Cepheids in Leavitt's catalogue, it is remarkably well-sampled, with 89 observations over the course of 45 cycles. Therefore, it is surprising that it is one of the Cepheids with the largest difference in $\log P$ with modern estimates. 

\begin{figure*}[h!]
\centering
\includegraphics[width=0.49\textwidth]{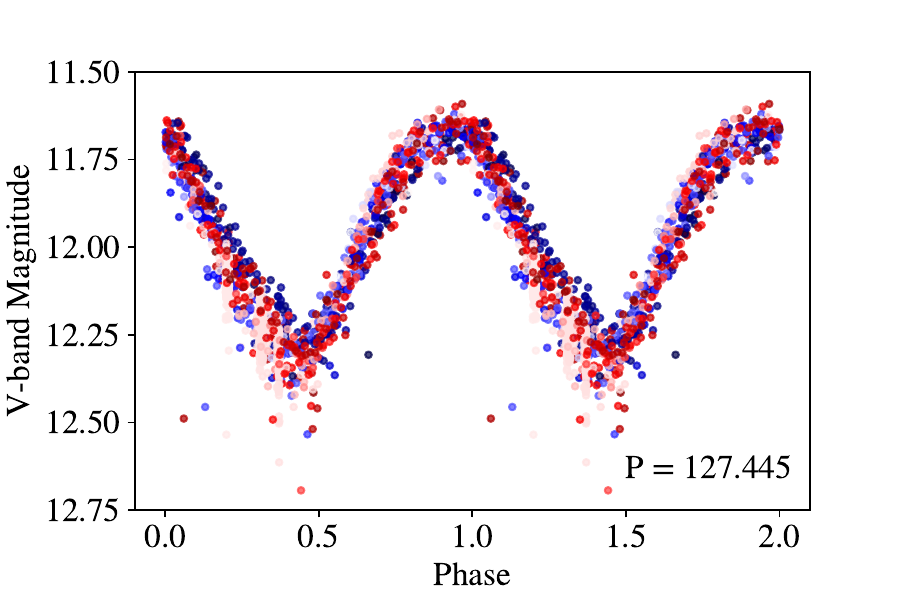} \includegraphics[width=0.49\textwidth]{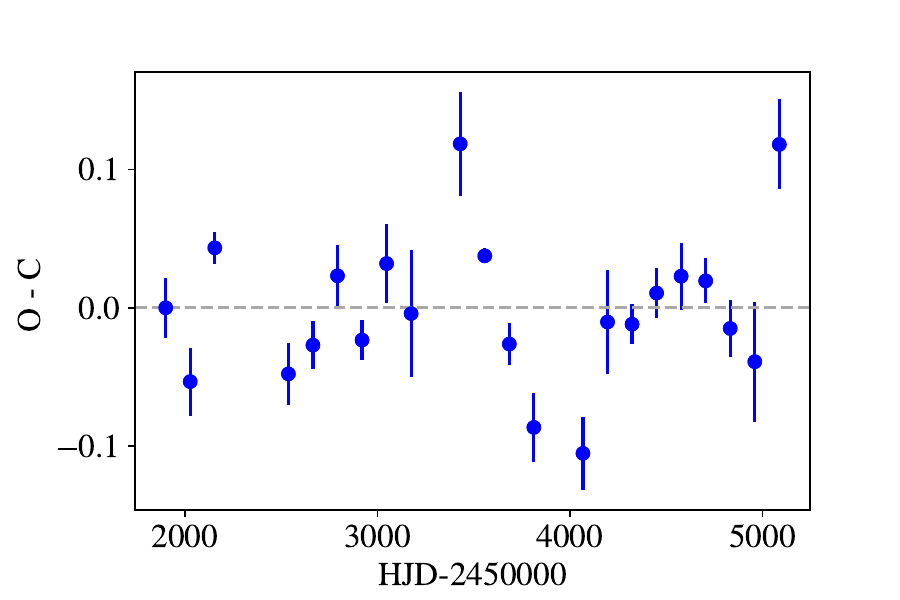} 
\caption{{\it Left:} Folded light curve for HV 821 (BZ Tuc) assuming a period of 127.445 days, which we find minimizes the phase dispersion during the span of the observations. Points are colored chronologically from blue (earlier) to red (later). {\it Right:} O-C diagram of HV 821 (BZ Tuc) constructed using ASAS time series collected over a baseline of 10 years. O-C is expressed as a fraction of the fit period. Uncertainties are estimated as $\sim \sigma_{phs}/\sqrt{N_{obs}}$ where $\sigma_{phs}$ is the phase uncertainty $N_{obs}$ is the number of observations per cycle. Only cycles observed near peak are included.}
\label{fig:PDM}
\end{figure*}

In order to investigate this period discrepancy, we collected ASAS light curves \citep{Pojmanski1997} covering about 10 years of observations, from 1998 to 2009. We folded these by the various literature periods and found that Leavitt's pulsation period minimizes the dispersion in phase slightly better than the OGLE period does during the span of the ASAS observations. We also determine a new period that minimizes the phase dispersion for this particular set of data ($P =$ 127.445 days). However, even with this period, the folded light curve still shows significant scatter from cycle-to-cycle (see Fig. \ref{fig:PDM}, left), which can be indicative of a period change. 

Using these observations, we constructed an O-C diagram for this Cepheid (Fig. \ref{fig:PDM}, right). Rather than a measurement error by Leavitt, the variations in the O-C diagram of HV 821 suggest that these different periods may be explained by an unstable period caused by a \textit{physical} period change, which is frequently observed in long-period Cepheids \citep[see][Table D]{Trahin2021}. Long-period Cepheids are expected to have more massive progenitors and to evolve more rapidly across the instability strip. The period instability of HV 821 was also briefly noted by \citet{Berdnikov_1997} who also speculated on its classification. Despite this, BZ Tuc has been classified by as a classical Cepheid in most modern catalogs. 

\newpage

\section{Henrietta Leavitt's Notebooks}
\label{appendixB}

\begin{figure*}[h!]
\centering
\includegraphics[width=0.90\textwidth]{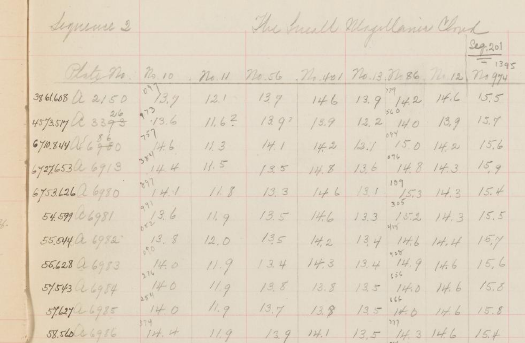} 
\caption{First few rows of page 25 of Henrietta Leavitt's notebook \citep{Leavitt_Notebook_1905}.}
\label{fig:notebook}
\end{figure*}

Figure \ref{fig:notebook} shows a few rows from page 25 of Henrietta Leavitt's notebook \citep{Leavitt_Notebook_1905}. The first column gives the observation date (equivalent of today's Julian Date), the second column is the plate number (e.g. A2150), and the next columns are the variable stars. For example, columns 3 and 4 are labeled "No. 10" and "No. 11", and correspond to HV 818 and HV 821 respectively. Magnitudes are listed in a "provisional scale" for each plate. These tables were used to produce the red light curves shown in Fig. \ref{fig:leavitt_light_curves}.  \\

\end{document}